\documentclass[12pt]{article}
\usepackage{array}
\usepackage[margin=1in]{geometry}
\usepackage{booktabs}
\usepackage{amsmath}
\usepackage{natbib}
\usepackage{graphicx, setspace, threeparttable, multicol, multirow, lscape}
\usepackage[dvipsnames]{xcolor}
\usepackage[
  colorlinks=true,
  linkcolor=red,
  citecolor=red,
  urlcolor=red,
  filecolor=red
]{hyperref}

\title{The Privilege of Exposure: Caste and Generative AI\\
       in India's Graduate Labour Market%
\thanks{[Acknowledgements. Funding. Disclaimer.]
The occupation-level exposure crosswalk and construction code are available at
\href{https://github.com/Kpmishra1998/nco2015-ai-exposure}%
     {\texttt{github.com/Kpmishra1998/nco2015-ai-exposure}}.}}

\author{Kaibalyapati Mishra%
\thanks{Institute for Social and Economic Change, Bengaluru.
Email: \href{mailto:kpmishra1998@gmail.com}{kpmishra1998@gmail.com}.}}

\date{\today}

\begin{document}
\maketitle

\begin{abstract}
\noindent
Who is exposed to generative AI in a developing-country labour market? We map
three occupational AI-exposure indices to India's redesigned Periodic Labour
Force Survey (2025) and document a steep caste gradient among 83,000 employed
graduates: graduates from Scheduled Castes and Scheduled Tribes are
0.24--0.37 standard deviations less exposed than upper-caste graduates within
the same district. Two channels drive the gap: one in four SC and one in
three ST graduates work in farm or elementary occupations untouched by AI,
and those in white-collar work are underrepresented in managerial, software,
and finance occupations. Because exposure commands a wage premium of up to 20
per cent, generative AI stands to widen, not narrow, India's caste earnings
gap.\\[4pt]
\end{abstract}

\pagebreak
\section{Introduction}

A growing literature measures which workers are exposed to artificial
intelligence by scoring the task content of occupations against AI
capabilities \citep{felten2021,eloundou2024,gmyrek2025}. Its consistent
finding is that exposure rises with education and is higher for women and
for white-collar workers \citep{pizzinelli2023,gmyrek2025}. What this
literature has not examined is how exposure is distributed across social
groups within a developing country, where occupational sorting reflects not
only skill but historical hierarchies of access. India is the natural case:
caste remains a first-order determinant of occupational attainment and wages
even among the highly educated \citep{madheswaran2007}, and the sectors
where generative AI adoption is most plausible---IT, finance, and business
services---are precisely those where upper-caste graduates are concentrated.

This paper provides, to our knowledge, the first caste-disaggregated
estimates of occupational exposure to AI. We construct a crosswalk linking
three independently built exposure indices to India's National Classification
of Occupations (NCO-2015) and merge it with the first wave of the redesigned
Periodic Labour Force Survey (PLFS) of 2025, covering roughly 83,000
employed graduates. Three results emerge. First, AI exposure among graduates
rises monotonically with caste advantage: relative to upper-caste
(``Others'') graduates in the same district, Scheduled Caste (SC) graduates
are 0.24--0.27 standard deviations less exposed and Scheduled Tribe (ST)
graduates 0.31--0.37 standard deviations less exposed, with the gap nearly
identical across all three indices. Second, the gap decomposes into two
channels: an occupational relegation margin---24.6 per cent of SC and 32.0
per cent of ST graduates work in farm or elementary occupations with
essentially zero AI exposure, against 12.1 per cent of Others---and sorting
within white-collar work, where upper-caste graduates are over-represented
in managerial, software, and finance occupations. Third, exposure of the
type associated with high-skill abilities carries a wage premium of roughly
20 log points per standard deviation, while the most automatable clerical
exposure does not: the relationship between automation-type exposure and
earnings is an inverted U.

Together these results invert the usual framing of AI exposure as
displacement risk. In India's graduate labour market, exposure is a marker
of privilege. SC and ST graduates are insulated from generative AI for the
worst possible reason: the labour market has already excluded them from the
occupations the technology will augment. A construction labourer with a
degree cannot be displaced by a language model; he has been displaced by
something older. If, as the wage results suggest, the gains from generative
AI accrue to workers in exposed, augmentable occupations, the technology
will compound rather than offset the caste earnings gap among graduates.

The paper relates to three literatures. It extends the cross-country
exposure work of \citet{pizzinelli2023} and \citet{gmyrek2025} to
within-country social stratification. It connects to evidence on AI-skill
demand in India, where vacancy data show AI demand concentrated in
high-wage urban services \citep{copestake2023}. And it adds a technological
dimension to the literature on caste gaps in graduate labour markets and
the devaluation of credentials under mass higher-education expansion
\citep{madheswaran2007}.

\section{Data and measurement}

\subsection{Labour force data}

We use the first-visit person-level file of the redesigned PLFS, covering
calendar year 2025. The redesigned survey moved to a monthly schedule with a
rotational panel; the first-visit file contains each sampled household once.
Our sample is all persons aged 15--65 who are employed by usual principal
activity status. Graduates are persons whose general education level is
graduate or postgraduate and above. This yields 449,041 employed persons, of
whom 82,830 are graduates. Social group is reported at the household level
in four categories: Scheduled Tribe, Scheduled Caste, Other Backward Classes
(OBC), and Others, the last serving as the reference category and consisting
predominantly of upper-caste households. All statistics use survey weights.
Table~\ref{tab:vardef} defines all variables and Table~\ref{tab:sumstats}
reports summary statistics.

\begin{table}[!htbp]\centering
\caption{Summary statistics}
\label{tab:sumstats}
\small
\begin{tabular}{lrrrrr}
\toprule
 & Mean & SD & Min & Max & $N$ \\
\midrule
\multicolumn{6}{l}{\emph{Panel A: All employed, aged 15--65}} \\
\addlinespace[2pt]
AIOE exposure ($z$)        & $-0.686$ & 0.881 & $-1.83$ & 1.68  & 448,801 \\
GPT exposure ($z$)         & $-0.532$ & 0.765 & $-1.48$ & 2.41  & 449,041 \\
ILO GenAI score ($z$)      & $-0.562$ & 0.835 & $-1.41$ & 2.78  & 449,041 \\
High-exposure share        & 0.038    & 0.171 & 0.00    & 1.00  & 449,041 \\
Age                        & 39.30    & 12.22 & 15      & 65    & 449,041 \\
Female                     & 0.307    & 0.461 & 0       & 1     & 449,041 \\
Urban                      & 0.297    & 0.457 & 0       & 1     & 449,041 \\
SC                         & 0.204    & 0.403 & 0       & 1     & 449,041 \\
ST                         & 0.109    & 0.312 & 0       & 1     & 449,041 \\
OBC                        & 0.458    & 0.498 & 0       & 1     & 449,041 \\
Farm/elementary occupation & 0.548    & 0.498 & 0       & 1     & 449,041 \\
Public sector              & 0.062    & 0.242 & 0       & 1     & 449,041 \\
Log monthly earnings       & 9.733    & 0.766 & 5.99    & 13.12 & 121,705 \\
\midrule
\multicolumn{6}{l}{\emph{Panel B: Graduates}} \\
\addlinespace[2pt]
AIOE exposure ($z$)        & 0.365    & 0.973 & $-1.83$ & 1.68  & 82,815 \\
GPT exposure ($z$)         & 0.323    & 0.861 & $-1.48$ & 2.41  & 82,830 \\
ILO GenAI score ($z$)      & 0.317    & 1.044 & $-1.41$ & 2.78  & 82,830 \\
High-exposure share        & 0.166    & 0.337 & 0.00    & 1.00  & 82,830 \\
Age                        & 35.66    & 10.37 & 19      & 65    & 82,830 \\
Female                     & 0.248    & 0.432 & 0       & 1     & 82,830 \\
Urban                      & 0.558    & 0.497 & 0       & 1     & 82,830 \\
SC                         & 0.124    & 0.330 & 0       & 1     & 82,830 \\
ST                         & 0.045    & 0.207 & 0       & 1     & 82,830 \\
OBC                        & 0.444    & 0.497 & 0       & 1     & 82,830 \\
Farm/elementary occupation & 0.180    & 0.384 & 0       & 1     & 82,830 \\
Public sector              & 0.196    & 0.397 & 0       & 1     & 82,830 \\
Log monthly earnings       & 10.238   & 0.755 & 6.11    & 13.12 & 45,765 \\
\bottomrule
\end{tabular}
\medskip
\begin{minipage}{0.9\textwidth}\footnotesize
\emph{Notes:} Means and standard deviations are survey-weighted; minima,
maxima, and observation counts are unweighted. Exposure scores are
standardised across the 130 NCO-2015 3-digit occupation groups, so sample
means reflect the occupational composition of employment. Log monthly
earnings computed for regular salaried workers with positive earnings.
Sample: employed by usual principal activity status, PLFS 2025 first visit.
\end{minipage}
\end{table}

\subsection{Exposure measurement}

PLFS records occupation at the 3-digit level of NCO-2015, which is
constructed to align with ISCO-08: at three digits the two classifications
correspond one to one. We attach three exposure indices to the 130 occupation
groups. The first is the AI Occupational Exposure index (AIOE) of
\citet{felten2021}, an ability-based measure built from O*NET. The second is
the task-based measure of \citet{eloundou2024}, the share of an occupation's
tasks for which a large language model with appropriate tooling can halve
completion time (their $\beta$, human-rated). Both are built on US
occupational data and reach NCO-2015 through the official Bureau of Labor
Statistics SOC--ISCO crosswalk, with scores aggregated within ISCO minor
groups using US employment weights. The third is the 2025 generative-AI
exposure index of the International Labour Organization \citep{gmyrek2025},
which is constructed directly on ISCO-08 task descriptions from Polish
task-level data, worker surveys, and expert assessment, and therefore
requires no US crosswalk. We aggregate its 4-digit scores to 3 digits by
unweighted means and additionally use its categorical exposure gradients.
All scores are standardised across the 130 occupation groups.

The obvious concern with transferring exposure scores across countries is
that the same occupation code may bundle different tasks in India than in
the United States \citep{carbonero2023}. Two features of our design address
this. The ILO index does not rely on US task data, and the three indices
agree closely: Spearman rank correlations across the 130 NCO groups are
0.92 between the two US-based measures and 0.81--0.83 between each of them
and the ILO index, almost exactly the range reported when the same indices
are compared on European data. Our results are nearly identical across all
three. The merge matches 100 per cent of weighted employment.

Throughout, exposure measures potential interaction between an occupation's
tasks and current AI capabilities, not adoption. We interpret the results as
the incidence of generative AI under counterfactual adoption, in the
standard tradition of this literature.

\section{Empirical strategy}

\subsection{The exposure gap}

Our baseline specification regresses standardised occupational AI exposure on social group indicators, conditioning on demographic composition and location:
\begin{equation}
E_i = \alpha + \beta_{SC}\,\text{SC}_i + \beta_{ST}\,\text{ST}_i + \beta_{OBC}\,\text{OBC}_i + X_i'\gamma + \delta_d + \varepsilon_i,
\label{eq:gap}
\end{equation}
where $E_i$ is one of the three standardised exposure scores (AIOE, GPT exposure, or ILO GenAI score) attached to graduate $i$'s 3-digit occupation, $X_i$ is a quadratic in age, sex, and a sector indicator, $\delta_d$ is a district fixed effect, and Others is the omitted social-group category. The regression is estimated by survey-weighted least squares; standard errors are clustered at the level at which exposure is assigned, the NCO-2015 3-digit occupation, since $E_i$ takes only 130 distinct values and is mechanically correlated within occupation. Table~\ref{tab:gaps} reports $\hat\beta_{SC}$, $\hat\beta_{ST}$, and $\hat\beta_{OBC}$ for each of the three indices.

\subsection{Decomposing the gap}

To trace the gap through occupational structure we estimate equation~\eqref{eq:gap} under four nested sets of controls. Column (1) includes no controls. Column (2) adds $X_i$ and $\delta_d$ as above. Column (3) further adds $R_i$, an indicator for employment in a farm or elementary occupation (NCO divisions 6 or 9), so that $\hat\beta_{SC}$ and $\hat\beta_{ST}$ net out the occupational-relegation margin. Column (4) replaces $R_i$ with a full set of NCO 1-digit occupation-division fixed effects $\delta_{o(i)}$, comparing graduates within the same broad occupation group. We label this an accounting decomposition rather than a causal design: occupation is the channel through which caste affects exposure, so conditioning on it apportions the raw gap across margins without identifying a caste effect net of occupation.

\subsection{Earnings and the exposure premium}

For regular salaried graduates with positive earnings we estimate
\begin{equation}
\ln w_i = \alpha + \rho\,E_i + X_i'\gamma + \delta_d + \varepsilon_i,
\label{eq:wage}
\end{equation}
with $X_i$ and $\delta_d$ as in equation~\eqref{eq:gap}. For the ILO GenAI score, whose categorical exposure gradients suggest a non-monotonic relationship with skill content, we additionally estimate the quadratic
\begin{equation}
\ln w_i = \alpha + \rho_1 E_i + \rho_2 E_i^2 + X_i'\gamma + \delta_d + \varepsilon_i.
\label{eq:wage_quad}
\end{equation}
To ask whether disadvantaged-group graduates capture less of the premium within exposed occupations, we interact exposure with social group,
\begin{equation}
\ln w_i = \alpha + \rho\,E_i + \sum_{g} \beta_g G_i^g + \sum_{g} \theta_g \left(E_i \times G_i^g\right) + X_i'\gamma + \delta_d + \varepsilon_i,
\label{eq:wage_interact}
\end{equation}
where $G_i^g \in \{\text{SC}_i, \text{ST}_i, \text{OBC}_i\}$; $\theta_{SC}$ and $\theta_{ST}$ are the coefficients of interest.

To trace the premium across the earnings distribution we estimate unconditional quantile regressions following the recentered influence function (RIF) approach, replacing $\ln w_i$ in equation~\eqref{eq:wage} with $\text{RIF}(\ln w_i; \tau)$ at each quantile $\tau$:
\begin{equation}
\text{RIF}(\ln w_i; \tau) = \alpha_\tau + \rho_\tau\,E_i + X_i'\gamma_\tau + \delta_d + \varepsilon_i,
\label{eq:rif}
\end{equation}
estimated separately for $\tau \in \{10, 25, 50, 75, 90\}$. The coefficient $\rho_\tau$ gives the marginal effect of a one standard deviation increase in exposure on the unconditional $\tau$-quantile of log earnings.

\subsection{Heterogeneity}

We examine whether the exposure gap compounds with gender by estimating equation~\eqref{eq:gap} separately by sex and, in a pooled specification, augmenting it with a female indicator and its interaction with each social-group indicator,
\begin{equation}
E_i = \alpha + \sum_{g} \beta_g G_i^g + \beta_F \text{Female}_i + \sum_{g} \theta_g \left(G_i^g \times \text{Female}_i\right) + X_i'\gamma + \delta_d + \varepsilon_i.
\label{eq:gender}
\end{equation}
We further re-estimate equation~\eqref{eq:gap} separately by age cohort (entrants, prime-age, senior) and by enterprise type (public sector versus private/other), and report robustness of $\hat\beta_{SC}$ and $\hat\beta_{ST}$ to unweighted exposure aggregation, exclusion of ICT occupations, and restriction to postgraduates.
\section{Results}

\subsection{The caste gradient in exposure}

Panel A of Table~\ref{tab:exposure} reports survey-weighted mean exposure of
employed graduates by social group. Exposure rises monotonically with caste
advantage on every index. Upper-caste graduates sit about half a standard
deviation above the occupation-level mean (0.50--0.57 depending on the
index); SC graduates sit essentially at the mean (0.07--0.08) and ST
graduates at or below it. The categorical measure tells the same story:
18.9 per cent of the occupational content of upper-caste graduate employment
falls in the ILO's two highest exposure gradients, against 12.5--13.6 per
cent for ST and SC graduates.

\begin{table}[!htbp]\centering
\caption{AI exposure of employed graduates by social group, PLFS 2025}
\label{tab:exposure}
\small
\begin{tabular}{lcccc}
\toprule
 & AIOE & GPT exposure & GenAI score & High-exposure \\
 & (Felten et al.) & (Eloundou et al.) & (ILO 2025) & share \\
\midrule
\multicolumn{5}{l}{\emph{Panel A: All employed graduates}} \\
\addlinespace[2pt]
SC      & 0.080 & 0.072 & 0.074    & 0.136 \\
ST      & 0.027 & 0.021 & $-$0.035 & 0.125 \\
OBC     & 0.303 & 0.268 & 0.264    & 0.158 \\
Others  & 0.567 & 0.502 & 0.496    & 0.189 \\
\midrule
\multicolumn{5}{l}{\emph{Panel B: Excluding farm and elementary occupations}} \\
\addlinespace[2pt]
SC      & 0.485 & 0.398 & 0.423 & \\
ST      & 0.517 & 0.415 & 0.381 & \\
OBC     & 0.619 & 0.530 & 0.552 & \\
Others  & 0.772 & 0.673 & 0.683 & \\
\midrule
\multicolumn{5}{l}{\emph{Panel C: Occupational relegation}} \\
\addlinespace[2pt]
 & \multicolumn{2}{c}{Farm/elementary share}
 & \multicolumn{2}{c}{$N$ (unweighted)} \\
\cmidrule(lr){2-3}\cmidrule(lr){4-5}
SC      & \multicolumn{2}{c}{0.246} & \multicolumn{2}{c}{10,024} \\
ST      & \multicolumn{2}{c}{0.320} & \multicolumn{2}{c}{6,638}  \\
OBC     & \multicolumn{2}{c}{0.199} & \multicolumn{2}{c}{35,606} \\
Others  & \multicolumn{2}{c}{0.121} & \multicolumn{2}{c}{30,562} \\
\bottomrule
\end{tabular}
\medskip
\begin{minipage}{0.92\textwidth}\footnotesize
\emph{Notes:} Survey-weighted means. Exposure indices are standardised
($z$-scores) across 130 NCO-2015 3-digit occupation groups. High-exposure
share is the weighted share of constituent ISCO-08 4-digit occupations in
ILO exposure Gradients 3--4. Panel C reports the weighted share of graduates
employed in farm (NCO division 6) or elementary (division 9) occupations.
Sample: employed persons aged 15--65 with general education level graduate
or above, usual principal activity status, PLFS 2025 first visit.
\end{minipage}
\end{table}

Table~\ref{tab:gaps} shows that the gradient is not an artefact of geography
or demographic composition. Within districts, conditional on a quadratic in
age, sex, and sector, SC graduates are 0.24--0.27 standard deviations less
exposed than upper-caste graduates, ST graduates 0.31--0.37 standard
deviations less exposed, and OBC graduates about 0.14 standard deviations
less exposed. The coefficients are remarkably stable across the three
indices---including the ILO index built without any US task data---and all
are significant at the 1 per cent level with standard errors clustered at
the level at which exposure is assigned.

\begin{table}[!htbp]\centering
\caption{Conditional AI-exposure gaps by social group}
\label{tab:gaps}
\small
\begin{tabular}{lccc}
\toprule
 & AIOE & GPT exposure & GenAI score \\
 & (Felten et al.) & (Eloundou et al.) & (ILO 2025) \\
 & (1) & (2) & (3) \\
\midrule
SC  & $-0.270^{***}$ & $-0.246^{***}$ & $-0.237^{***}$ \\
    & (0.078)        & (0.063)        & (0.070)        \\
\addlinespace
ST  & $-0.372^{***}$ & $-0.305^{***}$ & $-0.320^{***}$ \\
    & (0.064)        & (0.054)        & (0.063)        \\
\addlinespace
OBC & $-0.146^{***}$ & $-0.144^{***}$ & $-0.139^{***}$ \\
    & (0.032)        & (0.028)        & (0.028)        \\
\midrule
District FE   & Yes    & Yes    & Yes    \\
Controls      & Yes    & Yes    & Yes    \\
Observations  & 82,812 & 82,827 & 82,827 \\
$R^2$         & 0.216  & 0.221  & 0.193  \\
\bottomrule
\end{tabular}
\medskip
\begin{minipage}{0.80\textwidth}\footnotesize
\emph{Notes:} Dependent variables are standardised occupational AI-exposure
scores. Omitted category: Others. Controls: quadratic in age, sex, sector.
Survey-weighted least squares. Standard errors in parentheses, clustered by
NCO-2015 3-digit occupation.
$^{*}\,p<0.10$, $^{**}\,p<0.05$, $^{***}\,p<0.01$.
\end{minipage}
\end{table}

\subsection{Mechanisms: relegation and sorting}

Two channels generate the gradient. The first is occupational relegation.
Panel C of Table~\ref{tab:exposure} shows that 24.6 per cent of employed SC
graduates and 32.0 per cent of ST graduates work in farm or elementary
occupations---as cultivators, agricultural labourers, or construction
labourers---against 19.9 per cent of OBC and 12.1 per cent of upper-caste
graduates. These occupations have essentially zero AI exposure under every
index. The relegation margin is itself a measure of credential devaluation:
a substantial minority of disadvantaged-group degree holders work in
occupations that require no degree at all.

\begin{figure}[!htbp]\centering
\includegraphics[width=\textwidth]{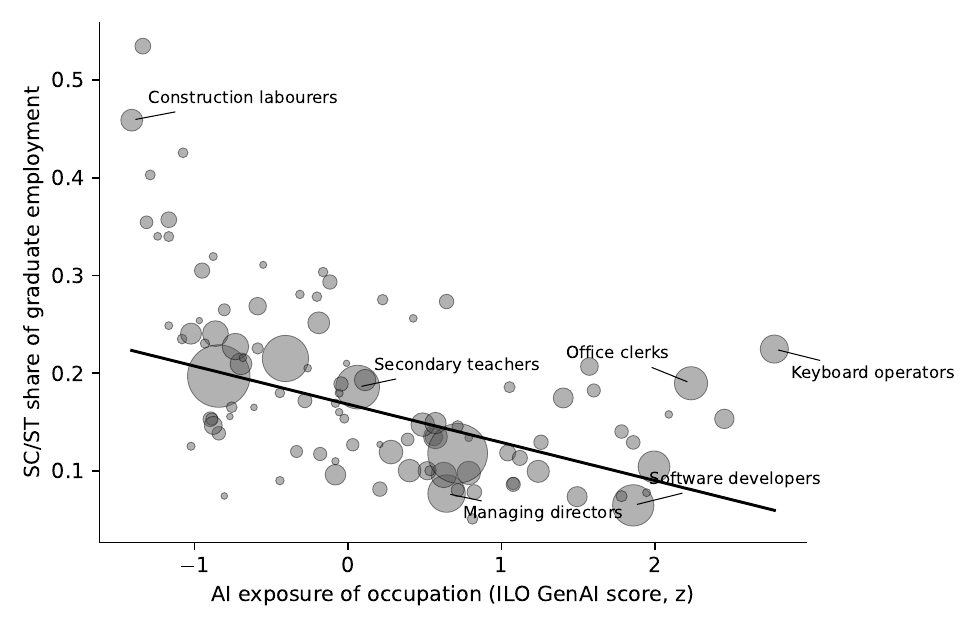}
\caption{Caste sorting across AI-exposed occupations}
\label{fig:sorting}
\medskip
\begin{minipage}{0.85\textwidth}\footnotesize
\emph{Notes:} Each circle is an NCO-2015 3-digit occupation group; the
horizontal axis is its standardised ILO GenAI exposure score and the
vertical axis the SC/ST share of graduate employment in the occupation
(survey-weighted). Circle area is proportional to weighted graduate
employment; occupations below 0.1 per cent of graduate employment are
omitted. The line is the employment-weighted linear fit. Sample: employed
graduates aged 15--65, PLFS 2025 first visit.
\end{minipage}
\end{figure}

The second channel is sorting within white-collar work. Panel B repeats
Panel A excluding farm and elementary occupations. The gap narrows by
roughly 40 per cent but does not close: among white-collar graduates,
upper-caste workers remain 0.26--0.30 standard deviations more exposed than
SC workers. The occupational composition behind this residual is stark.
Among upper-caste graduates, 5.3 per cent are software developers, 3.4 per
cent finance professionals, and 9.4 per cent hold managerial occupations,
including 5.1 per cent recorded as managing directors and chief executives.
Among SC graduates the corresponding software and finance shares are 2.1
and 1.9 per cent, and no managerial occupation appears among their twelve
largest occupation groups. Teaching, by contrast, absorbs a similar share
of both groups (about 11 per cent), so public-sector teaching employment is
not what differentiates them; exclusion from exposed private white-collar
occupations is. Figure~\ref{fig:sorting} summarises the sorting at the
occupation level: the SC/ST share of graduate employment declines steadily
with exposure, from nearly half in construction labour to under a tenth in
software development.

\subsection{Why exposure matters: the wage premium}

Table~\ref{tab:wages} asks whether exposure is an asset or a liability by
estimating earnings regressions for regular salaried graduates. A one
standard deviation increase in ability-based exposure (AIOE) is associated
with 20.0 log points higher monthly earnings; the task-based measure of
\citet{eloundou2024} gives 18.4 log points. The ILO automation score, by
contrast, carries no linear premium (column 3), and column 4 shows why: its
relationship with earnings is an inverted U, rising over most of the
distribution and turning down at the top of the automation scale, where the
most automatable clerical occupations sit. This is the familiar
routine-cognitive pattern \citep{autor2013}: occupations exposed through
high-skill abilities are well paid, while occupations exposed through
routine automatable tasks are not.

\begin{table}[!htbp]\centering
\caption{AI exposure and graduate earnings}
\label{tab:wages}
\small
\begin{tabular}{lcccc}
\toprule
 & \multicolumn{4}{c}{Log monthly earnings} \\
\cmidrule(lr){2-5}
 & (1) & (2) & (3) & (4) \\
\midrule
AIOE (Felten et al.)           & $0.200^{***}$ &               &         &               \\
                               & (0.036)       &               &         &               \\
\addlinespace
GPT exposure (Eloundou et al.) &               & $0.184^{***}$ &         &               \\
                               &               & (0.043)       &         &               \\
\addlinespace
GenAI score (ILO 2025)         &               &               & 0.047   & $0.189^{***}$ \\
                               &               &               & (0.043) & (0.064)       \\
\addlinespace
GenAI score squared            &               &               &         & $-0.090^{**}$ \\
                               &               &               &         & (0.035)       \\
\midrule
District FE   & Yes    & Yes    & Yes    & Yes    \\
Controls      & Yes    & Yes    & Yes    & Yes    \\
Observations  & 45,757 & 45,759 & 45,759 & 45,759 \\
$R^2$         & 0.359  & 0.348  & 0.320  & 0.335  \\
\bottomrule
\end{tabular}
\medskip
\begin{minipage}{0.85\textwidth}\footnotesize
\emph{Notes:} Sample: regular salaried graduates with positive monthly
earnings. Exposure scores standardised as in Table~\ref{tab:gaps}. Controls:
quadratic in age, sex, sector. District fixed effects. Survey-weighted least
squares; standard errors in parentheses, clustered by NCO-2015 3-digit
occupation. $^{*}\,p<0.10$, $^{**}\,p<0.05$, $^{***}\,p<0.01$.
\end{minipage}
\end{table}

The wage structure gives the caste gradient its economic content. The
exposure that SC and ST graduates lack is precisely the kind that pays.
Interacting exposure with social group yields negative coefficients for SC
($-0.043$, $p = 0.09$) and ST ($-0.078$, $p = 0.05$), suggestive evidence
that disadvantaged-group graduates also capture less of the premium within
exposed occupations, though we do not lean on these marginal estimates. The
robust facts are the sorting results: disadvantaged-group graduates are
systematically absent from the occupations where generative AI meets high
wages.

\subsection{Decomposing the gap}

Table~\ref{tab:decomp} traces the exposure gap through a sequence of nested
specifications. The raw gap between SC and upper-caste graduates is 0.42
standard deviations and the ST gap 0.53 (column 1). Demographic composition
and location absorb a little under half of each (column 2). Adding an
indicator for farm or elementary employment (column 3) shows the channels
differ by group: relegation accounts for over half of the remaining ST gap,
which falls from 0.32 to 0.15, but only a sixth of the SC gap. Column 4
replaces the indicator with NCO division (1-digit) fixed effects, comparing
graduates within broad occupation groups. Even there, SC graduates are 0.13
and ST graduates 0.16 standard deviations less exposed: roughly half of the
conditional gap operates \emph{within} occupation divisions---between, say,
teaching and software development inside the professional division---rather
than between manual and white-collar work. We stress that columns (3) and
(4) are an accounting decomposition, not a causal design: occupation is the
channel through which exposure is determined, so these columns apportion the
gap across margins rather than identify the effect of caste net of
occupation.

\begin{table}[!htbp]\centering
\caption{Decomposing the exposure gap (ILO GenAI score)}
\label{tab:decomp}
\small
\begin{tabular}{lcccc}
\toprule
 & Raw & +Dem., district FE & +Relegation & +Occ.\ division FE \\
 & (1) & (2) & (3) & (4) \\
\midrule
SC  & $-0.422^{***}$ & $-0.237^{***}$ & $-0.201^{***}$ & $-0.127^{***}$ \\
    & (0.092)        & (0.070)        & (0.047)        & (0.028)        \\
\addlinespace
ST  & $-0.531^{***}$ & $-0.320^{***}$ & $-0.153^{**}$  & $-0.163^{***}$ \\
    & (0.111)        & (0.063)        & (0.064)        & (0.049)        \\
\addlinespace
OBC & $-0.232^{***}$ & $-0.139^{***}$ & $-0.111^{***}$ & $-0.058^{***}$ \\
    & (0.048)        & (0.028)        & (0.027)        & (0.017)        \\
\addlinespace
\midrule
Observations & 82,830 & 82,827 & 82,827 & 82,827 \\
$R^2$        & 0.024  & 0.193  & 0.363  & 0.700  \\
\bottomrule
\end{tabular}
\medskip
\begin{minipage}{0.9\textwidth}\footnotesize
\emph{Notes:} Column (1): no controls. Column (2) adds quadratic in age,
sex, sector, and district FE. Column (3) adds an indicator for farm or
elementary occupation. Column (4) replaces it with NCO 1-digit division FE.
Survey-weighted; SEs clustered by NCO-2015 3-digit occupation. Conditioning
on occupation is a decomposition, not a causal control.
$^{*}\,p<0.10$, $^{**}\,p<0.05$, $^{***}\,p<0.01$.
\end{minipage}
\end{table}

\subsection{Heterogeneity: gender, cohorts, and the public sector}

Table~\ref{tab:gender} examines the intersection with gender. The caste
gradient is present within both sexes, somewhat larger among men (SC gap
$-0.26$ against $-0.17$ among women), and the caste--gender interactions in
the pooled specification are positive but individually insignificant: caste
and gender disadvantage do not compound. The female main effect is itself
noteworthy. Among upper-caste graduates, women are 0.21 standard deviations
\emph{less} exposed than men---the opposite of the cross-country pattern, in
which women's concentration in clerical work makes them more exposed
\citep{pizzinelli2023,gmyrek2025}. Among Indian graduates the dominant
female occupations are teaching and health, which every index scores as
unexposed, so the gender sorting that protects women from automation also
excludes them from augmentable work.

\begin{table}[!htbp]\centering
\caption{Exposure gaps by caste and gender (ILO GenAI score)}
\label{tab:gender}
\small
\begin{tabular}{lccc}
\toprule
 & Men & Women & Pooled \\
 & (1) & (2) & (3) \\
\midrule
SC  & $-0.258^{***}$ & $-0.166^{***}$ & $-0.259^{***}$ \\
    & (0.079)        & (0.058)        & (0.075)        \\
\addlinespace
ST  & $-0.350^{***}$ & $-0.224^{***}$ & $-0.339^{***}$ \\
    & (0.065)        & (0.068)        & (0.067)        \\
\addlinespace
OBC & $-0.150^{***}$ & $-0.094^{***}$ & $-0.155^{***}$ \\
    & (0.031)        & (0.035)        & (0.030)        \\
\addlinespace
Female            &         &         & $-0.212^{**}$ \\
                  &         &         & (0.100)       \\
\addlinespace
SC $\times$ Female  &       &         & $0.093$       \\
                    &       &         & (0.058)       \\
\addlinespace
ST $\times$ Female  &       &         & $0.076$       \\
                    &       &         & (0.076)       \\
\addlinespace
OBC $\times$ Female &       &         & $0.069$       \\
                    &       &         & (0.045)       \\
\addlinespace
\midrule
Observations & 62,713 & 20,098 & 82,827 \\
$R^2$        & 0.189  & 0.274  & 0.194  \\
\bottomrule
\end{tabular}
\medskip
\begin{minipage}{0.9\textwidth}\footnotesize
\emph{Notes:} Columns (1)--(2): estimated separately by sex. Column (3):
pooled with interactions. Controls and FE as in Table~\ref{tab:gaps}.
$^{*}\,p<0.10$, $^{**}\,p<0.05$, $^{***}\,p<0.01$.
\end{minipage}
\end{table}

Table~\ref{tab:cohort} splits the sample by age. The SC gap is largest among
labour-market entrants aged 15--30 ($-0.28$) and declines monotonically
across cohorts ($-0.22$ prime, $-0.19$ senior); the OBC gap falls from
$-0.18$ to $-0.06$. The exposure divide is therefore not a legacy of older
cohorts working its way out of the labour force: it is widest precisely
where generative AI is arriving, among graduates entering employment now.
Only the ST gap is flat across cohorts, consistent with relegation---its
dominant channel---being persistent across generations.

\begin{table}[!htbp]\centering
\caption{Exposure gaps by age cohort (ILO GenAI score)}
\label{tab:cohort}
\small
\begin{tabular}{lccc}
\toprule
 & Entrants (15--30) & Prime (31--45) & Senior (46--65) \\
 & (1) & (2) & (3) \\
\midrule
SC  & $-0.282^{***}$ & $-0.216^{***}$ & $-0.191^{***}$ \\
    & (0.078)        & (0.063)        & (0.068)        \\
\addlinespace
ST  & $-0.330^{***}$ & $-0.333^{***}$ & $-0.300^{***}$ \\
    & (0.087)        & (0.083)        & (0.080)        \\
\addlinespace
OBC & $-0.175^{***}$ & $-0.141^{***}$ & $-0.063^{**}$  \\
    & (0.035)        & (0.030)        & (0.029)        \\
\addlinespace
\midrule
Observations & 31,972 & 34,538 & 16,272 \\
$R^2$        & 0.260  & 0.183  & 0.217  \\
\bottomrule
\end{tabular}
\medskip
\begin{minipage}{0.9\textwidth}\footnotesize
\emph{Notes:} Each column restricts the graduate sample to the stated age
band. Controls: age, sex, sector; district FE; SEs clustered by occupation.
$^{*}\,p<0.10$, $^{**}\,p<0.05$, $^{***}\,p<0.01$.
\end{minipage}
\end{table}

Table~\ref{tab:public} splits employment by enterprise type. Inside the
public sector the caste gaps compress sharply (SC $-0.14$, ST $-0.14$);
outside it they widen (SC $-0.27$, ST $-0.44$). The contrast is starkest
for ST graduates, whose private-sector gap is more than three times the
public-sector gap. This is what one would expect if reservation compresses
occupational sorting within government employment while private hiring
reproduces it. The implication cuts both ways: the public sector is where
disadvantaged-group graduates come closest to occupational parity, but
public-sector occupations are concentrated in the unexposed range, so
parity is achieved in precisely the segment that generative AI will leave
behind.

\begin{table}[!htbp]\centering
\caption{Exposure gaps inside and outside the public sector}
\label{tab:public}
\small
\begin{tabular}{lcc}
\toprule
 & Public sector & Private/other \\
 & (1) & (2) \\
\midrule
SC  & $-0.140^{***}$ & $-0.266^{***}$ \\
    & (0.036)        & (0.078)        \\
\addlinespace
ST  & $-0.137^{**}$  & $-0.444^{***}$ \\
    & (0.065)        & (0.058)        \\
\addlinespace
OBC & $-0.107^{***}$ & $-0.146^{***}$ \\
    & (0.033)        & (0.031)        \\
\addlinespace
\midrule
Observations & 19,256 & 63,560 \\
$R^2$        & 0.140  & 0.247  \\
\bottomrule
\end{tabular}
\medskip
\begin{minipage}{0.8\textwidth}\footnotesize
\emph{Notes:} Public sector defined from enterprise type of principal
activity. Dependent variable: standardised ILO GenAI score. Controls:
quadratic in age, sex, sector; district FE; survey-weighted; SEs clustered
by NCO-2015 3-digit occupation.
$^{*}\,p<0.10$, $^{**}\,p<0.05$, $^{***}\,p<0.01$.
\end{minipage}
\end{table}

\subsection{The exposure premium across the earnings distribution}

\begin{figure}[!htbp]\centering
\includegraphics[width=\textwidth]{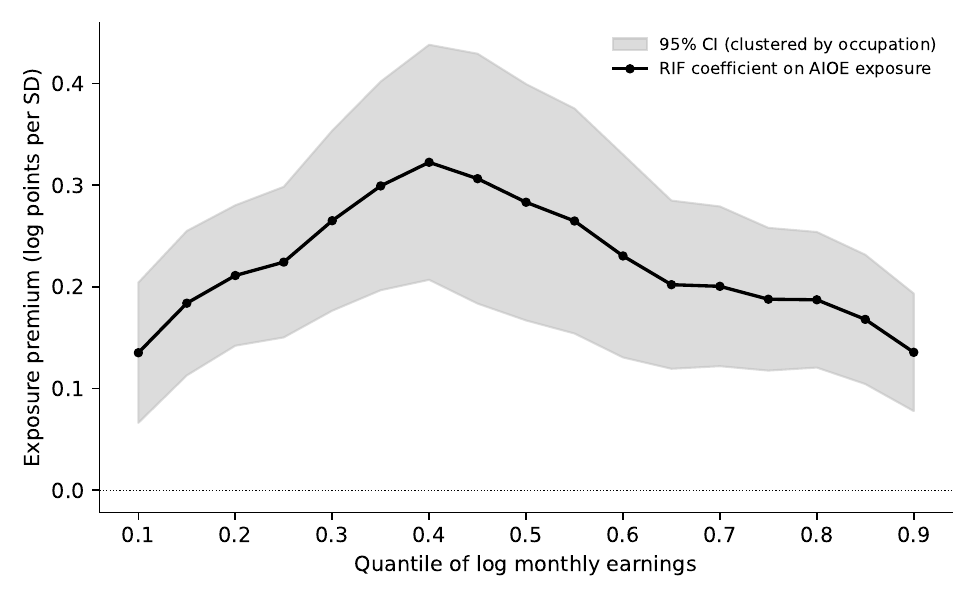}
\caption{The AI-exposure premium across the earnings distribution}
\label{fig:premium}
\medskip
\begin{minipage}{0.85\textwidth}\footnotesize
\emph{Notes:} Coefficients on standardised AIOE exposure from unconditional
quantile (RIF) regressions of log monthly earnings, estimated at every fifth
percentile from the 10th to the 90th. Controls: quadratic in age, sex,
sector; district fixed effects; survey-weighted. Shaded band: 95 per cent
confidence interval, standard errors clustered by NCO-2015 3-digit
occupation. Sample: regular salaried graduates with positive earnings.
\end{minipage}
\end{figure}

Table~\ref{tab:rif} estimates unconditional quantile (RIF) regressions of
log earnings on AIOE exposure. The premium is positive and significant at
every quantile but follows a hump: 13.5 log points at the 10th percentile,
rising to a peak of roughly 32 log points around the 40th percentile before
declining to about 14 at the 90th (Figure~\ref{fig:premium}). Exposed
occupations most strongly differentiate earnings in the middle of the
graduate wage distribution, where clerical and junior professional work
sits; at the top, high earnings are attainable in both exposed (software,
finance) and unexposed (medicine, senior management) occupations. The
distributional reading reinforces the main result: for the typical salaried
graduate, holding an AI-exposed occupation is associated with substantially
higher pay, and it is exactly this middle-of-the-distribution premium from
which SC and ST graduates are disproportionately absent.

\begin{table}[!htbp]\centering
\caption{Exposure premium across the earnings distribution (RIF regressions)}
\label{tab:rif}
\small
\begin{tabular}{lccccc}
\toprule
 & $q_{10}$ & $q_{25}$ & $q_{50}$ & $q_{75}$ & $q_{90}$ \\
 & (1) & (2) & (3) & (4) & (5) \\
\midrule
AIOE exposure & $0.135^{***}$ & $0.224^{***}$ & $0.283^{***}$
              & $0.188^{***}$ & $0.136^{***}$ \\
              & (0.035)       & (0.037)       & (0.059)
              & (0.035)       & (0.029)       \\
\addlinespace
\midrule
Observations & 45,757 & 45,757 & 45,757 & 45,757 & 45,757 \\
$R^2$        & 0.190  & 0.260  & 0.263  & 0.219  & 0.144  \\
\bottomrule
\end{tabular}
\medskip
\begin{minipage}{0.9\textwidth}\footnotesize
\emph{Notes:} Unconditional quantile (RIF) regressions of log monthly
earnings on standardised AIOE exposure. Sample, controls, FE, and clustering
as in Table~\ref{tab:wages}.
$^{*}\,p<0.10$, $^{**}\,p<0.05$, $^{***}\,p<0.01$.
\end{minipage}
\end{table}

\subsection{Robustness}

Table~\ref{tab:robust} re-estimates the conditional gaps under alternative
measurement choices: unweighted aggregation of exposure scores within ISCO
groups, exclusion of ICT occupations---where the assumption that Indian and
US task content coincide is weakest---and restriction to postgraduates. The
SC and ST coefficients move little. The postgraduate column is the only one
with notable attenuation (SC $-0.16$, ST $-0.25$), suggesting the gradient
is somewhat flatter at the highest education level, though it remains
significant at 1 per cent throughout.

\begin{table}[!htbp]\centering
\caption{Robustness of the conditional exposure gap}
\label{tab:robust}
\small
\begin{tabular}{lcccc}
\toprule
 & Baseline & Unweighted agg. & Excl.\ ICT & PG only \\
 & (1) & (2) & (3) & (4) \\
\midrule
SC  & $-0.237^{***}$ & $-0.251^{***}$ & $-0.206^{***}$ & $-0.156^{***}$ \\
    & (0.070)        & (0.079)        & (0.065)        & (0.049)        \\
\addlinespace
ST  & $-0.320^{***}$ & $-0.365^{***}$ & $-0.289^{***}$ & $-0.245^{***}$ \\
    & (0.063)        & (0.065)        & (0.058)        & (0.079)        \\
\addlinespace
OBC & $-0.139^{***}$ & $-0.148^{***}$ & $-0.122^{***}$ & $-0.121^{***}$ \\
    & (0.028)        & (0.031)        & (0.025)        & (0.027)        \\
\addlinespace
\midrule
Observations & 82,827 & 82,812 & 79,653 & 20,644 \\
$R^2$        & 0.193  & 0.238  & 0.158  & 0.220  \\
\bottomrule
\end{tabular}
\medskip
\begin{minipage}{0.9\textwidth}\footnotesize
\emph{Notes:} Column (2) uses unweighted within-ISCO aggregation of the
AIOE. Column (3) drops ICT occupations (NCO 133, 251, 252, 351, 352).
Column (4) restricts to postgraduates. Controls and FE as in
Table~\ref{tab:gaps}.
$^{*}\,p<0.10$, $^{**}\,p<0.05$, $^{***}\,p<0.01$.
\end{minipage}
\end{table}

\section{Conclusion}

In India's graduate labour market, exposure to generative AI is a privilege
gradient. Upper-caste graduates dominate the managerial, software, and
finance occupations that AI will augment; SC and ST graduates are
concentrated in unexposed work, a quarter to a third of them in farm and
elementary occupations that require no degree. Their insulation from AI is
the footprint of their prior exclusion. Since exposure of the augmentable
kind commands a substantial wage premium, the distribution of generative
AI's gains will follow the distribution of exposure, and the technology
should be expected to widen the caste earnings gap among graduates rather
than narrow it.

The policy implication is not to slow adoption but to widen access to the
occupations and skills through which AI's gains flow. Demand for AI skills
in India is growing rapidly and is concentrated in high-wage urban services;
whether disadvantaged-group graduates can enter those occupations will
determine whether the technology entrenches or erodes existing hierarchies.

Three caveats bound the interpretation. Exposure is not adoption, and
adoption in India outside IT and finance remains limited; our estimates
describe incidence under counterfactual adoption. Task content may differ
between Indian and US occupations, though the agreement between US-based and
ISCO-native indices limits the force of this concern. And 3-digit
occupational coding averages over heterogeneous detailed occupations, which
attenuates rather than inflates the gradients we report.

\begin{landscape}
\begin{table}[!htbp]\centering
\caption{Variable definitions and sources}
\label{tab:vardef}
\small
\begin{tabular}{p{4.0cm}p{9cm}p{3.6cm}}
\toprule
Variable & Definition & Source \\
\midrule
AIOE exposure & Ability-based AI occupational exposure index, standardised
($z$) across 130 NCO-2015 3-digit groups; US employment-weighted aggregation
within ISCO minor groups & Felten et al.\ (2021); own crosswalk \\
\addlinespace
GPT exposure & Share of occupational tasks for which an LLM with tooling
halves completion time (human-rated $\beta$), standardised as above &
Eloundou et al.\ (2024); own crosswalk \\
\addlinespace
ILO GenAI score & Mean generative-AI automation score, ISCO-08 4-digit,
aggregated to 3 digits by unweighted mean, standardised as above &
Gmyrek et al.\ (2025) \\
\addlinespace
High-exposure share & Share of constituent ISCO-08 4-digit occupations in
ILO exposure Gradients 3--4 & Gmyrek et al.\ (2025) \\
\addlinespace
SC, ST, OBC & Household social group: Scheduled Caste, Scheduled Tribe,
Other Backward Classes; omitted category Others &
PLFS household schedule \\
\addlinespace
Farm/elementary occupation & Usual principal occupation in NCO division 6
(skilled agricultural) or 9 (elementary occupations) & PLFS; NCO-2015 \\
\addlinespace
Public sector & Enterprise type of principal activity: government/local
body, public sector enterprise, or autonomous body &
PLFS person schedule \\
\addlinespace
Graduate & General education level graduate or postgraduate and above &
PLFS person schedule \\
\addlinespace
Log monthly earnings & Log of monthly earnings from regular salaried
employment, positive earners only & PLFS person schedule \\
\addlinespace
Survey weight & PLFS multiplier divided by 100 & PLFS \\
\bottomrule
\end{tabular}
\end{table}
\end{landscape}

\end{document}